\documentclass[a4paper]{jpconf}
\usepackage{graphicx}

\usepackage{bbm}
\usepackage[utf8]{inputenc}
\usepackage{placeins}
\usepackage{amssymb,amsmath,amsfonts}
\usepackage[colorlinks,linkcolor=purple,citecolor=teal]{hyperref}
\usepackage{dsfont}
\usepackage{color}
\usepackage{hyphenat}
\usepackage{wrapfig}
\usepackage{empheq}
\usepackage{textcomp}
\usepackage[caption=false]{subfig}
\usepackage{rotating}
\usepackage{wrapfig}
\usepackage{pdfpages}
\usepackage{verbatim}
\usepackage{setspace}
\usepackage{array}
\usepackage{caption}
\usepackage[pdf]{pstricks}
\usepackage{upgreek}
\usepackage{cmll}
\usepackage{latexsym}
\usepackage{braket}
\usepackage{cite}
\usepackage{epsfig}
\usepackage[left=2.4cm,top=3.3cm,right=2.4cm,bottom=3.3cm,bindingoffset=0cm]{geometry}
\usepackage[titletoc,page]{appendix}
\usepackage{multicol}
\usepackage{youngtab}
\newcommand{\beq}{\begin{eqnarray}}
\newcommand{\eeq}{\end{eqnarray}}
\newcommand{\bea}{\begin{eqnarray}}
\newcommand{\eea}{\end{eqnarray}}
\newcommand{\be}{\begin{equation}}
\newcommand{\ee}{\end{equation}}
\def\brc{\langle}
\def\ckt{\rangle}

\def\Tr{\qopname\relax o{Tr}}

\numberwithin{equation}{section}

\begin{document}
\title{Quantum fluctuations, particles and  entanglement: 
solving  the quantum measurement problems 
\footnote{Talk presented in  DICE2022,
September 19-23, 2022 (Castiglioncello, Tuscany).
To appear in  DICE2022 Proceedings, 
Journ. Phys.: Conf. Series".  }
}

\author{Kenichi Konishi}

%

\address{ 

INFN, Sezione di Pisa,    Largo Pontecorvo, 3, Ed. C, 56127 Pisa, Italy,  \\[2pt]

Department of Physics ``E. Fermi", University of Pisa,  \\[2pt]

Largo Pontecorvo, 3, Ed. C, 56127 Pisa, Italy    \\[2pt]

}

\ead{kenichi.konishi@unipi.it}

\begin{abstract}
 The so-called quantum measurement problems are solved from a new perspective. 
One of the main  observations is that the basic entities of our world are  {\it  particles}, elementary or composite. 
 It follows that  each elementary process, hence each measurement process at its core, is a spacetime,  pointlike, event. 
  Another key idea is that, when a  microsystem  $\psi$  gets into contact with the experimental device,  factorization of $\psi$  rapidly fails and entangled mixed states appear.
 The wave functions  for the microsystem-apparatus coupled system for different measurement outcomes then lack overlapping spacetime support. 
It means that  the aftermath of each measurement  is a single term in the sum:  a ``wave-function collapse". 
Our  discussion leading  to a diagonal density matrix, $\rho= {\rm diag} ( |c_1|^2, \ldots, |c_n|^2, \ldots )$ shows how 
the information encoded  in the wave function    $|\psi\ckt    =    \sum_n  c_n   | n \ckt$  gets transcribed,  via entanglement with the experimental device and environment, into the relative frequencies   ${\cal P}_n = |c_n|^2$  for various experimental outcomes $F=f_n$.   
Our discussion  represents the first, significant steps  towards filling in the logical gaps in the conventional 
interpretation based on Born's rule, replacing it with a clearer understanding of quantum mechanics.
 Accepting objective reality of quantum fluctuations, independent of any experiments, and independently of human presence,  one renounces
  the idea that in a fundamental, complete theory of Nature 
the result of each  single experiment must necessarily be predictable.

\end{abstract}

\newpage

\section{Quantum measurement problems}

Quantum mechanics is described by the following few formulas,
\bea   &&    [x_i,p_j]= i \hbar \, \delta_{ij}  \;, \qquad  \Delta x \Delta p \ge    \frac{\hbar}{2}\;;   \nonumber \\
&&    i \hbar \frac{d \psi}{dt} =   H \psi\;; \nonumber \\
&&  H \psi_n = E_n \psi_n\,, \qquad H=  \frac{{\bf p}^2}{2m} + V\;, \qquad   {\bf p}= - i \hbar \nabla\;. 
\eea
The rest is, essentially,  just details.   In spite of the fundamental  remarkable simplicities, quantum mechanics (QM)
has led to 

 \begin{description}
  \item[(i)]  An amazing success and detailed confirmation in atomic physics,  ushering us into a century of impressive advance in basic scientific knowledge
  and  modern technologies, {\it  unprecedented in human history.}   We still live in the wake of this ongoing revolution.
  
  \item[(ii)]  Further pursuit into subatomic physics has eventually led to the standard model of the fundamental interactions \cite{Weinberg}-\cite{GellMann}, a non-Abelian gauge theory based on the gauge group,  
 \be    SU(3)_{QCD}\times  \{  SU(2)_L\times U(1)\}_{GWS}, 
 \ee
 (GWS stands for Glashow-Weinberg-Salam, and QCD for  Quantum Chromodynamics), describing very accurately the strong (nuclear) interactions and 
  electroweak interactions, down to the distance scales of order of $O(10^{-18})$ $cm$. 
  
  \item[(iii)]   One cannot forget about the beautiful phenomena in condensed matter physics, such as 
  superconductivity, quantum Hall effect, Bose-Einstein condensation of cold atoms, etc. etc. 
  
\end{description}

We can certainly talk about

\begin{center}    {\bf  $100$  years of extraordinary scientific achievements}.    \end{center}

\noindent  But, at the same time,  the QM predictions are, supposedly,  given by the probabilistic (Born's) rule. And this has  led  to  

\begin{center}  {\bf  $100$ years of uneasy feeling and doubts},     \end{center}

\noindent  that something fundamental is missing in our understanding   of QM.    These are often expressed as various (apparent) puzzles   or conceptual  difficulties, which are all loosely called the   ``quantum measurement problems" \cite{WheelerZ}-\cite{Bell}.    They are actually a collection of different questions:

\begin{description}
  \item[(a)]      Wave function collapse?
  \item[(b)]   Macroscopic superposition of states?  
  \item[(c)]   Forever branching manyworlds tree? 
    \item[(d)]   Spontaneous collapse of micro state, of the experimental device, and of the world? 
      \item[(e)]   Wave function just a bookkeeping device? 
        \item[(f)]   Quantum jumps vs Schr\"odinger equation? 
          \item[(g)]   Born's rule from the Schr\"odinger equation involving everything? 
            \item[(h)]   Quntum nonlocality, Hidden variables? 
                        \item[(i)]   How is it possible that the fundamental, complete theory of physics is incapable of unique prediction? 
\end{description}
The following discussion, based on \cite {KK},   will address all of them.  

\begin{figure}
\begin{center}
\includegraphics[width=6in]{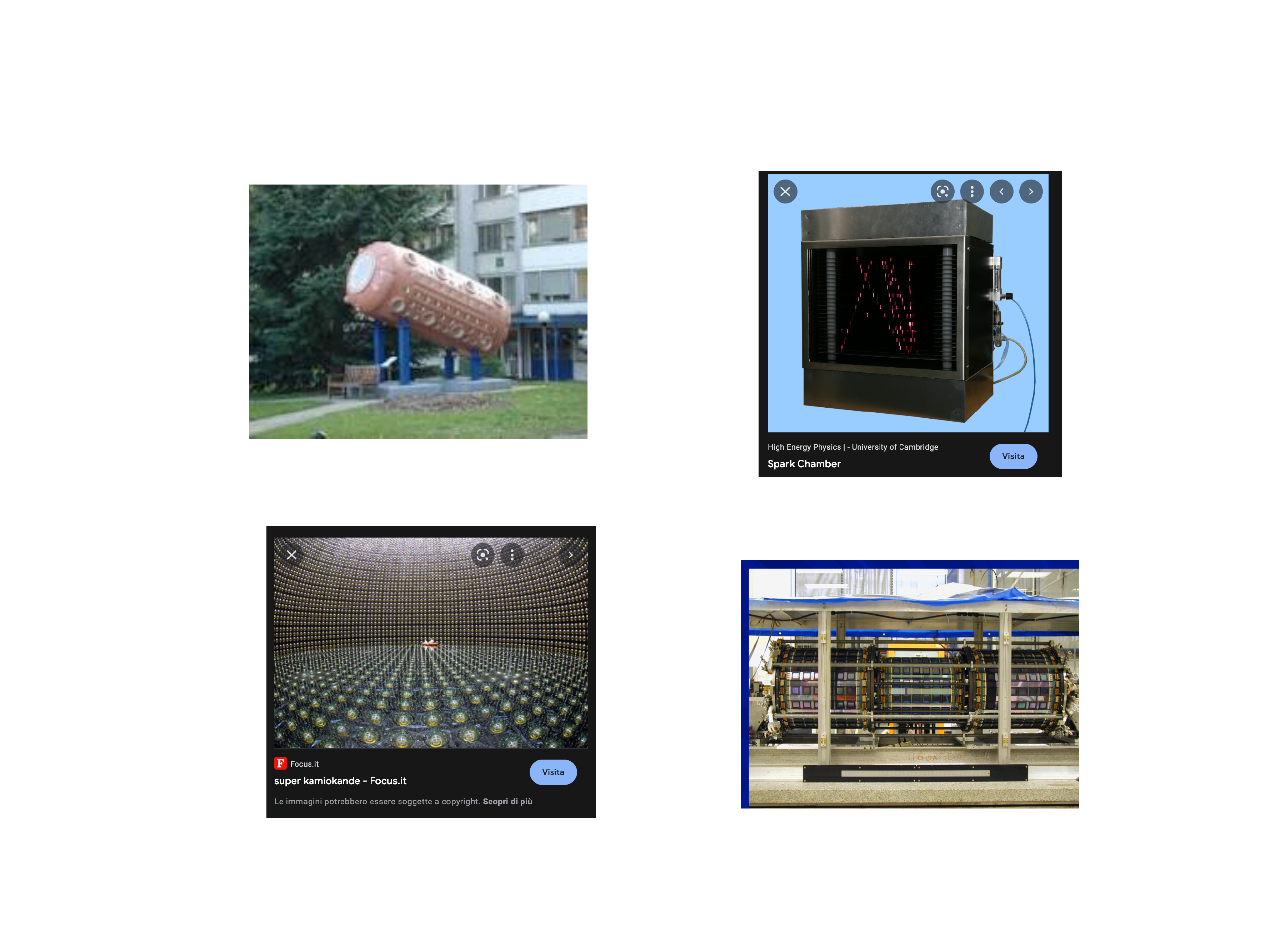}
\caption{\footnotesize   A few examples of quantum measurement devices.  }
\label{FigAppr}
\end{center}
\end{figure}

The actual  measurement devices can vastly differ in their nature, size, materials used, and technologies employed,  from a simple photographic plate, cloud and bubble  chambers filled with liquid or vapour, spark chambers and MWPC made 
with metal plates, wires and gas,  the neutrino detectors made of a huge tank of pure water and tens of thousands of  photomultipliers,  to the state-of-the-art  silicon detectors and some future apparatus which uses superconducting materials for dark matter search.   See Fig.~\ref{FigAppr}.
 Such an enormous diversity of the experimental devices requires  that  an equally vast simplification be made, to capture the essence of   quantum measurements.   
 
Indeed,  independently of the details,  
a good experimental device  faithfully reflects the quantum fluctuations of the system, described by the wave function $|\psi\ckt$.  Indeed, the result of each single experiment is, in general, apparently random and unpredictable  \footnote{An exception occurs when the state $\psi$
 is  one of the eigenstates of the quantity $F$, with $F= f_m$,  i.e.,   $|\psi\ckt = |m \ckt$.   In such a case, a good experiment  produces the same result $F=f_m$  every time.
  }.    
   Yet,  the information encoded  in the wave function $|\psi\ckt$
manifests itself  as the expectation values  for any variable  \footnote{Throughout, we use simply  the word ``variable",  for  a dynamical variable, physical quantity,   such as energy, momentum, angular momentum, position, etc., which is measurable  (known as an observable). Also, we use the same symbol $F$ for the physical variable itself and for the self-adjoint operator  representing it.  
The alert reader will not have any difficulty  telling which is meant, each time.
 }    $F$ (and all functions thereof),   
    \be       {\bar F}  \equiv     \brc  \psi | F  | \psi \ckt \;,     \label{fund}  
 \ee
that is,  as a (frequency-)  average  of the experimental outcomes,    $  {\bar  {f_n }}$,      
 \be    {\bar {f_n}} =   \sum_n       f_n  \,  {\cal P}_n   \;, \qquad    {\cal P}_n  =     |\brc  n  | \psi \ckt|^2 \;,  \label{notBorn}  
 \ee
 where  $F |n\ckt = f_n | n \ckt$.     ${\cal P}_n$ is  the relative frequency  for  $f_n$.  
 Such a  prediction of quantum mechanics is verified 
  by countless experiments.   That is,  ``QM works". 
  
The prediction (\ref{notBorn}) follows,  if one assumes  that  the  {\it  probability } of  each single measurement of $F$ 
  in  the state $\psi$ to give $f_n$,     is given by  
$    P_n  =     |\brc  n  | \psi \ckt|^2 \;,     \label{Born}   $
that is, by  Born's rule.   Born's rule   is presented in most textbooks as one of the fundamental postulates  of quantum mechanics 
\footnote{A rare exception is the famous book by  Dirac  (1958)  \cite{Dirac},  
     where slightly difference nuance is used.   }.    

  So what {\it is} the problem?   Well,  ``the quantum measurement problems", {\bf (a) }  $\sim$  {\bf  (i)},    above!

Throughout, we will   denote the  {\it  (normalized) relative frequency} with the calligraphy font  ${\cal P}_n$,    and write 
the  standard {\it  probability}  by using  the normal font  $P_n$.    Although they express mathematically the same numbers, they differ conceptually, and
lead to distinct logical conclusions.  

\section{Three aspects of QM }

Before analyzing the measurement processes  themselves  in Sec.~\ref{measurement},  we  must first discuss briefly  three familiar aspects of  QM. Though singly well known,  clarifying their precise roles in the measurement processes will be essential in the following  discussions. 

\subsection{Expectations values versus  Born's rule}

To fix the idea,  we consider the system described by the wave function,
\be  |\psi \ckt  = \sum_n   c_n\, | n \ckt\;,  \qquad   F  | n \ckt = f_n  | n \ckt\;, \qquad  \sum_n |c_n|^2 =1\;, 
\ee
where $F$ is the variable we are interested in. 
Repeated, identical measurement of $F$ will give an average,   (\ref{notBorn}).

Now,   (\ref{notBorn})  certainly follows from Born's rule,  (\ref{Born}).   In other words, the latter is a sufficient condition for the  former.    The problem is:  
is Born's rule also necessary?   

The answer is yes,   if  the following equality holds
\be    {\bar{O( f_n)}} \equiv  \sum_n {\cal P}_n   O(f_n)   =       \brc  \psi | O(F)  |\psi\ckt   \equiv    \sum_n  |c_n|^2   O(f_n)\;,  
\ee
 for general function  $O(F)$ of $F$. Then it can be shown that   
 \be      {\cal P}_n  =    |c_n|^2 =  P_n\;.  
 \ee
 The  proof  and illustrations are  given in \cite{KK}.

\subsection{Pure and mixed states}

A system described by a wave function (or a state vector)  $\psi$  
is a pure state.  It contains the complete knowledge of the system.

Otherwise,  the system is in a mixed state (or a mixture).  It is described by the density matrix, $\rho$.

For instance,  a subsystem   $A \subset \Sigma$ of a closed system  $\Sigma$, described by the wave function, 
\be    |\Psi \ckt =   \sum_{n, \alpha}   c_{n, \alpha}  |n\ckt | \alpha\ckt   \;, \qquad   \sum_{n, \alpha}  |^2 = 1\;,
\ee
the expectation value of  $G$ (pertinent to  the subsystem $A$) is given by  
\be  {\bar   G}=   \brc \Psi | G | \Psi \ckt  =  \Tr  {\mathbbm G} \rho\;,  \qquad     {\mathbbm G}_{mn}   \equiv  \brc m | G | n \ckt\;,  
\ee 
where the density matrix is given in this case  by 
\be  \rho_{n m}    \equiv   \sum_{\alpha}   c_{n, \alpha}  c_{m, \alpha}^*   \;.   \label{subset}
\ee

More generally the density matrix represents any sort of ignorance about the system, and will have forms different from  (\ref{subset}), but the trace formula
\be    {\bar   G}=   \Tr  {\mathbbm G} \rho\;,
\ee
is valid always.

\subsection{Factorization vis-\`a-vis entanglement     \label{factent} }

The question is, symbolically,  this: how can we study a single hydrogen atom, an electron, etc.,  when we know that the wave function involving identical fermions must be antisymmetrized (Fermi-Dirac (FD) statistics)?   

The answer is the lack of the spatial support in the wrong component.  For instance,  for an electron in the laboratory and another in the Sun, 
the wave function will look like
\be  \Psi=   \frac{1}{\sqrt{2}}   \big(\psi_{lab}( {\bf r}_1)   \psi_{sun}({\bf r}_2)-    \psi_{lab}({\bf r}_2)   \psi_{sun}({\bf r}_1)  \Big)   \sim    \psi_{lab}( {\bf r}_1)\;,   
\ee
for  
\be   {\bf r}_1  \in Lab.\;, \qquad    {\bf r}_2     \in Sun
\ee
as clearly  there is no support for the wrong component,  $\psi_{lab}( {\bf r}_2)  $ or  $  \psi_{sun}( {\bf r}_1)$.   

One sees   how a pure state emerges as the result of factorization. 
\begin{description}
  \item[(i)]   If the anstisymmetrization  is in the spin state, instead of the position states, then the wrong component is negligible, but  $\psi_{lab}( {\bf r}_1)$ 
  is in a spin mixed state;
  \item[(ii)]   If one takes two hydrogen atoms (two electrons)  $\sim  2 \,m$ apart, in    the next room in the laboratory, instead of the second in the Sun, 
  the factorization failure  (the effect of the wrong component) is still  very small, 
  \be   \sim 10^{-18}\;, 
  \ee
  as it follows by  using Bohr's radius!  And this gives some idea about the exactness of  some of our statement below,   even if some of the  
  argument about the spacetime localization below  concerns {\it spacetime support} of the wave functions, rather than the spatial  support, considered here.

  \item[(iii)]  Actually, in spite of the impression this example might have given to the readers,   the FD or BE   statistics are not fundamental for the purpose of explaining entanglement.  {\it  Any systems}  which have interacted in the past are  entangled, and in general, in a mixed state.   
  
\end{description}

We  note that factorization and  entanglement  are two faces of the same medal,  both characerizing quantum mechanics universally. 
Note in particular  that 
\begin{itemize}
\item     {\bf Factorization} works remarkably strongly;  it makes QM a workable, sensible, physics theory, in spite of the fact that everything is in principle
entangled in our universe.    It is   {\bf factorization}  that makes  the  pure state  a significant  concept, both from  theoretical and experimental points of view;

\item  {\bf Entanglement}    leads to the fascinating, characteristic phenomena in QM, such as quantum nonlocality, violation of Bell's   \cite{Bell} and CHSH \cite{CHSH}  inequalities, quantum cryptgraphy, etc.   At the same time, the uncontrolled or uncontrollable entanglement with the unobserved systems, is responsible for   docoherence, and 
emergence of classical physics  \cite{Zurek}-\cite{Zurek2}.

\end{itemize}

\noindent {\it  \bf  A   reflection}

In principle, we always live in a mixed state:  everything is interrelated, and interacted or interacting with each other,  in our universe.    
It is the notion of a pure quantum state, described by a wave function $\psi$,   which is extraordinary, and exceptional.  
{\it They arise as a result of factorization.}

These pure states are carefully prepared by experimentalists in  many small bubbles in the world, called physics laboratories,  equipped  with a good clean room and sofiscated  (and expensive)  vacuum and cryogenic technologies.

But they occur also naturally,   on the earth,  in the form of radioactivity,  which produces $\alpha$ ($\alpha$ particles), $\beta$  (electrons)  or $\gamma$  (photons) rays.  They also fall from the sky, in the form of cosmic rays   (energetic protons,  pions, etc.).    Then there was light.   Light, as it turns out, is a gas of non-interacting, thus, free, photons. They are all pure quantum states.

In a hindsight,   we see  why  these phenomena  (providing  pure quantum states  for free!)  have led to the discovery of quantum physics by  Planck, at the dawn of the 20th century.

\section{The measurement processes    
 \label{measurement}  }

Now we come to our main problem: the quantum measurement processes.

\subsection{Effective spacetime localization -  state-vector reduction   \label{reduction}}

The measurement of the variable $F$ in the state
\be  |\psi\ckt =   \sum_n  c_n \, | n \ckt  \;,   \qquad F |n\ckt = f_n  | n\ckt\;, 
\ee
is often assumed to proceed as 
 \bea     |\psi \ckt \otimes |\Phi_0\ckt  \otimes   | Env_0 \ckt  &=&   \left( \sum_n  c_n  | n \ckt  \right)    \otimes |\Phi_0\ckt  \otimes    | Env_0 \ckt   \label{step0}  \\
   & \longrightarrow &      \left( \sum_n  c_n  | n \ckt     \otimes |\Phi_n\ckt  \right )   \otimes   | Env_0 \ckt      \label{step1} \\
  & \longrightarrow &     \sum_n  c_n  | n \ckt     \otimes |\Phi_n\ckt    \otimes   | Env_n \ckt   \;,\label{step2} 
 \eea
 where in the first step  (\ref{step1})  the experimental device reads the measurement results,  $f_n$, and in  the final step   (\ref{step2})  the ``environment"  comes be aware of that result (the experimentalist has seen  the 
 result on her/his  computer screen).  
 These formulas appear to suggest a coherent superposition of distinct macroscopic states   
 leading to paradoxes, debates and endless confusion (the notorious example of this being Schr\"odinger's cat  ``paradox", see below).

 Actually,  the factorized form in  (\ref{step0})-(\ref{step2}), with various   $\otimes$  symbols,  is not correct,   except for the  factorization of  $ |\psi \ckt $ in the first line,  namely, before the experiment.   
 
We note that the factorized form  $|\Phi\ckt  \otimes   | Env \ckt $  is incorrect,   even before  the measurement.   The air molecules,  the casing of the device, etc., all show  that     $|\Phi\ckt  $ and   $  | Env \ckt $  are actually entangled.  Not only,   the boundaries between   $|\Phi\ckt  $ and   $  | Env \ckt $ are
badly defined, and  to be regarded as just a convention.    Nevertheless, any experimentalist knows what the relevant part of the device is.  She (or he) knows
perfectly well that there is no need to push the boundary between the measuring device  $|\Phi\ckt  $ and  the rest of the world  $  | Env \ckt $
up to inside the human brain,   in order to ensure good, precise measurement results.  

Keeping this state of matter in mind,   we denote the  measurement-device-environment entangled ``state", as 
\be     |\Phi;  Env\ckt\;,
\ee
below.  The measurement process can be represented more appropriately as   \footnote{The device-enviroment ``state", $ |\Phi_0;    Env_0 \ckt $,  
can never be identical,  at  two different measurement instants,  see   discussions below (\ref{recover}).    }
       \bea     |\psi \ckt \otimes |\Phi_0;  Env_0 \ckt  &=&   \left( \sum_n  c_n  | n \ckt  \right)    \otimes   |\Phi_0;    Env_0 \ckt   \label{step000}  \\
   & \longrightarrow &       \sum_n  c_n  \,    |{\tilde  \Phi}_n;     Env_0 \ckt      \label{step111} \\
  & \longrightarrow &     \sum_n  c_n   \,   |{\tilde  \Phi}_n;     Env_n \ckt       \;.\label{step222} 
 \eea
 where  $ {\tilde  \Phi}_n $ stands for  the entangled state of the microsystem-apparatus with the reading of the measurement result, $f_n$.  The exact timing 
 of passage from  the first stage of  measurement-registering of the result on the apparatus (\ref{step111})  to the  second (\ref{step222})   (e.g.,  the moment  in which  the experimentalist sees the  
result  on her  or his computer screen; others read about it in Physical Review),   
 is entirely immaterial.

 Emergence of the mixed state and the state-vector reduction  (the ``wave function collapse")  occurs as follows. 
 
 {\it  Just before the measurement},   the system  is described by a wave function of factorized form by assumption, 
\be    |\Psi\ckt_{t=0_-}  =      \left( \sum_n  c_n  | n \ckt  \right)    \otimes   |\Phi_0;    Env_0 \ckt  \;.
\ee
The expectation value of any  generic  variable  $G$ in  this state (prior to the measurement) is given by  
\be   {\bar G}=    \brc \Psi | G  | \Psi\ckt   =       \Tr       {\mathbbm G} \rho \;, \qquad  ( {\mathbbm G} )_{mn} \equiv    \brc m | G | n  \ckt\;, 
\ee
with the  density matrix  having the form  
\be  \rho=    \left(\begin{array}{ccccc}|c_1|^2  &  c_1 c_2*   &  c_1 c_3^*  &   \ldots  &  \ldots  \\   c_2 c_1^*    & |c_2|^2 &  c_2 c_3^* & \ldots   &  \\   
c_3 c_1^*  &  & \ddots &  &  \\  \vdots    &  &  & |c_n|^2 &  \\ &  &  &  & \ddots\end{array}\right)\;,   \label{pure}
\ee 
characteristic of a pure state, i.e.,   $\rho_{nm} =  c_n c_m^*$.    The normalization condition 
\be          \brc  \Phi_0;    Env_0   | \Phi_0;    Env_0 \ckt  =1\;
\ee
has been used. 
In the case of the particular variable  $F$  (whose eigenstates are $|n\ckt$'s),  $  {\mathbbm F} $  is diagonal,  and its  quantum average is given by  the known formula 
\be   {\bar F} =    \Tr  ( {\mathbbm F} \rho )  = \sum_n   f_n  |c_n|^2\;. \label{usual}  
   \ee
   
{\it  As soon as the microsystem gets into contact with the experimental device, and the measurement events (a chain ionization process, hadronic cascade, etc.) have taken place},  the total wave function takes an entangled form, (\ref{step111})  
\be    |\Psi\ckt_{t=0_+}  =     \sum_n  c_n      |{\tilde  \Phi}_n;     Env_0 \ckt  \;.    \label{inspite}
\ee
$ {\tilde  \Phi}_n $ denotes  the entangled microsystem-apparatus  state with the reading of the measurement result, $f_n$.

\begin{figure}
\begin{center}
\includegraphics[width=5 in]{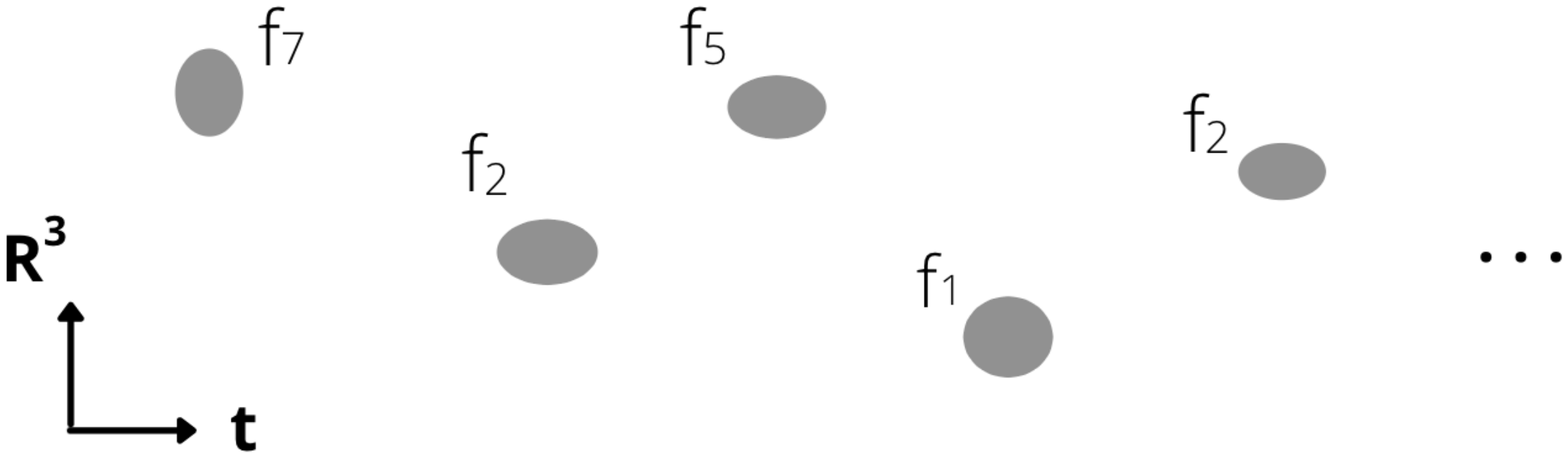}
\caption{\footnotesize Measurement of a variable $F$.  Each blob describes a single measurement event (a chain ionization reaction, hadronic cascade, etc.) occurring at a localized region in spacetime, 
with the experimental results,  $F=f_7, f_2, f_5, f_1, f_2$, etc.  The wave functions describing the  different measurement events ${\tilde  \Phi}$ have no overlapping spacetime supports, therefore are orthogonal.
}
\label{mixture}
\end{center}
\end{figure}

Now, the wave function describing   ${\tilde  \Phi}_m$  (the aftermath of a measurement, with the result, $F=f_m$) and  that for    ${\tilde  \Phi}_{n}$
 (the aftermath of a measurement, with the result, $F=f_{n}$),  corresponding to two  distinct spacetime events,   have no overlapping 
 spacetime support, as illustrated in Fig.~\ref{mixture}.   (\ref{inspite})   is a mixed state.  
 
 Treating the formula (\ref{inspite}) as a mixed state, and not a coherent superposition of states, eliminates at once the contradiction found in a recent paper \cite{Renner}.   The argument of \cite{Renner} may be regarded as an independent  confirmation of the fact  that the state after the measurement, (\ref{inspite}),  cannot be a coherent superposition \footnote{A  possible issue in \cite{Renner}  could  be 
the fact that the information transmission from the first laboratory to the second  about the result of the measurement made in the first, might render invalid the assumption that the two laboratories are isolated
quantum systems. 
}.

   The  expectation values of a variable  $G$ are then  given by
   \be   {\bar G}=    \brc \Psi | G  | \Psi\ckt   =       \sum_{m,n}  c_m^* c_n  \brc  {\tilde  \Phi}_m;     Env_0  | \, G \,  |   {\tilde  \Phi}_n;     Env_0 \ckt   \;;
   \ee
   but the fact    ${\tilde  \Phi}_m$  and ${\tilde  \Phi}_{n}$   ($m \ne n$)     have no common spacetime support,
     means that  {\it   for any local operator } $G$,   orthogonality relations, 
       \be      \brc    {\tilde  \Phi}_m;     Env_0     |   \, G \, |     {\tilde  \Phi}_{n};     Env_0 \ckt =0\;, \qquad  m \ne n  \;,  \label{lack}  
\ee
   hold.     Consequently,   $  {\bar G} $ is given by  the sum of the diagonal terms
   \be    {\bar G} =    \sum_n   |c_n|^2  G_{nn} \;,    \qquad   G_{nn}=    \brc    {\tilde  \Phi}_n;     Env_0     |  \,G \,|    {\tilde  \Phi}_{n};     Env_0 \ckt \;.     \label{above} 
   \ee  
This     means  that the density matrix has been  reduced to a  diagonal form,   
\be  \rho=    \left(\begin{array}{ccccc}|c_1|^2  &     &    &  &  \\    & |c_2|^2 &  &  &  \\      &  & \ddots &  &  \\  &  &  & |c_n|^2 &  \\ &  &  &  & \ddots\end{array}\right)\;.\label{diagonal}
\ee 
For the particular case of the variable  $F$,   we recover the standard prediction, 
\be    {\bar F} =    \sum_n   |c_n|^2   f_n \;, \qquad  .^.. \quad    {\cal P}_n=    |c_n|^2\;, \label{recover} 
\ee
where $ {\cal P}_n$ are the normalized relative frequencies for different outcomes  $f_n$.
    
   The emergence of the mixed state   (\ref{diagonal})  
is often attributed to the fact that  a typical  experimental device is made of a macroscopic  (difficult-to-specify)  number of atoms and molecules, and it is not possible to keep track of the phase relations among different terms  in (\ref{step111}) to any significant extent.

Also,  the macroscopic device  $\Phi$, evolving in $t$ entangled with $Env$, can never be in an identical quantum state 
$ |\Phi_0; Env_0 \ckt $   at  two different  measurement instants.  This is in stark contrast with  the  {\it  identical quantum state } for the microsystem, $|\psi\ckt$, which can be  and is indeed produced via e.g., a repeatable experiment (state preparation - see Sec.~\ref{WFcollapse})    for each  measurement. 

The observations above are both  certainly correct, but  we need another,
 crucial ingredient for  the decoherence in the measurement processes:   the lack  of the common  spacetime supports  in the wave functions,  and the consequent orthogonality and decoherence among  terms corresponding to different measurement results,  (\ref{lack}).  Importance of this is that   {\it   it implies  that 
 the result of each measurement event  is a state-vector reduction,   
\be    \left( \sum_n  c_n  | n \ckt  \right)    \otimes |\Phi_0; Env \ckt    \Longrightarrow     |{\tilde \Phi}_m ; Env \ckt \;:      \label{JumpCor}
 \ee
 i.e.,  with a single term present,  the instant after the measurement (e.g., with $F=f_m$).    
 This state of affair is perceived by us as a  ``wave-function collapse".  }  
 
 Summarizing:
 \begin{description}
  \item[A:]    The spacetime eventlike nature of the triggering  particle-measuring-device interactions  introduces an effective spacetime localization of each measurement event; 
  \item[B:]   At the moment  the microsystem - a pure quantum state $|\psi\ckt$ - gets into contact with the measurement device-environment state $ |\Phi_0; Env_0 \ckt $  which is a mixed state,
     factorization of $|\psi\ckt$ gets lost rapidly, and an entangled, mixed (classical) state   $ |\Phi_n; Env_0 \ckt$ with the unique recording,  $F= f_n$,  is generated.  
\end{description}
The result is the state-vector reduction,  (\ref{JumpCor}).

\begin{figure}
\begin{center}
\includegraphics[width=2.4in]{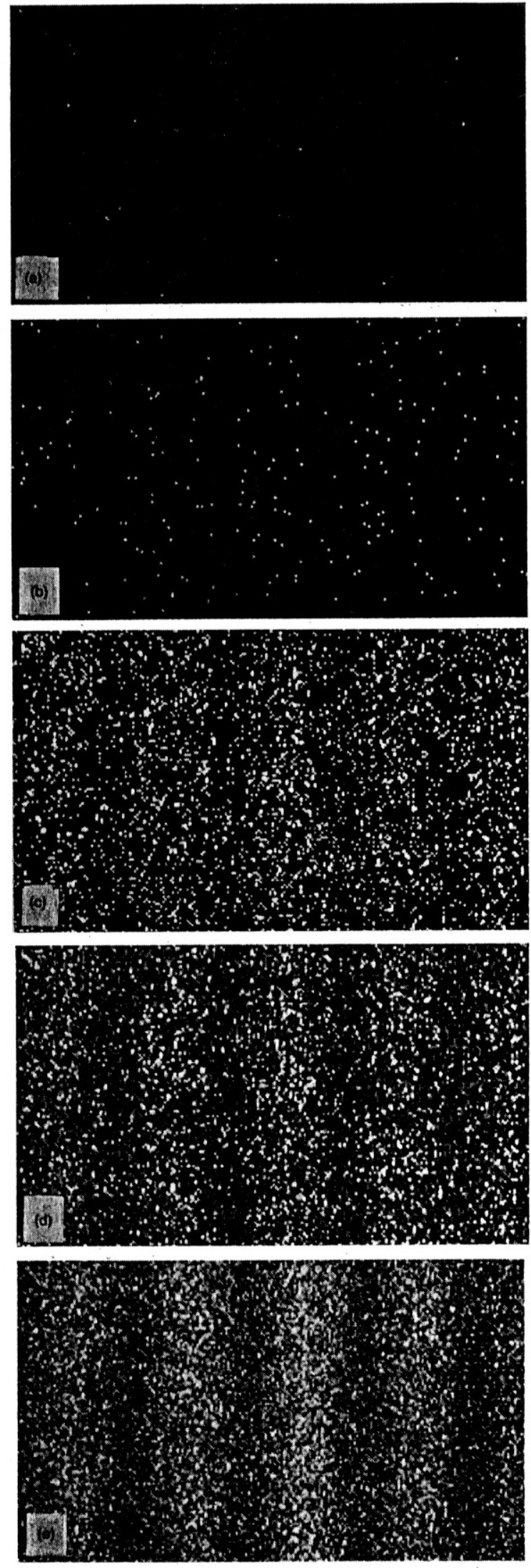}
\caption{\small  The pictures correspond, from the top to the bottom,  to successive  exposures to $10$, $100$, $3000$,  $20000$ and $70000$ electrons, respectively.
Reproduced from   \cite{Tonomura},    A~. Tonomura, J~. Endo, T~. Matsuda, T~. Kawasaki and H.~Ezawa,
``Dimonstration of single-electron buildup of interference pattern",
  American Journal of Physics  57, 117 (1989); doi: 10.1119/1.16104, 
 with the permission of the American Association of Physics Teachers.
}
\label{TonomuraExp}
\end{center}
\end{figure}

The effective spacetime localization and  apparent wave-function collapse, can be nicely visualized in the ``double-slit"   experiments by Tonomura et. al, \cite{Tonomura}, see Fig.~\ref{TonomuraExp}.

 \subsection{Repeatable,  nonrepeatable  and intermediate types of  state reductions
   \label{WFcollapse} }

\begin{figure}
\begin{center}
\includegraphics[width=4.5in]{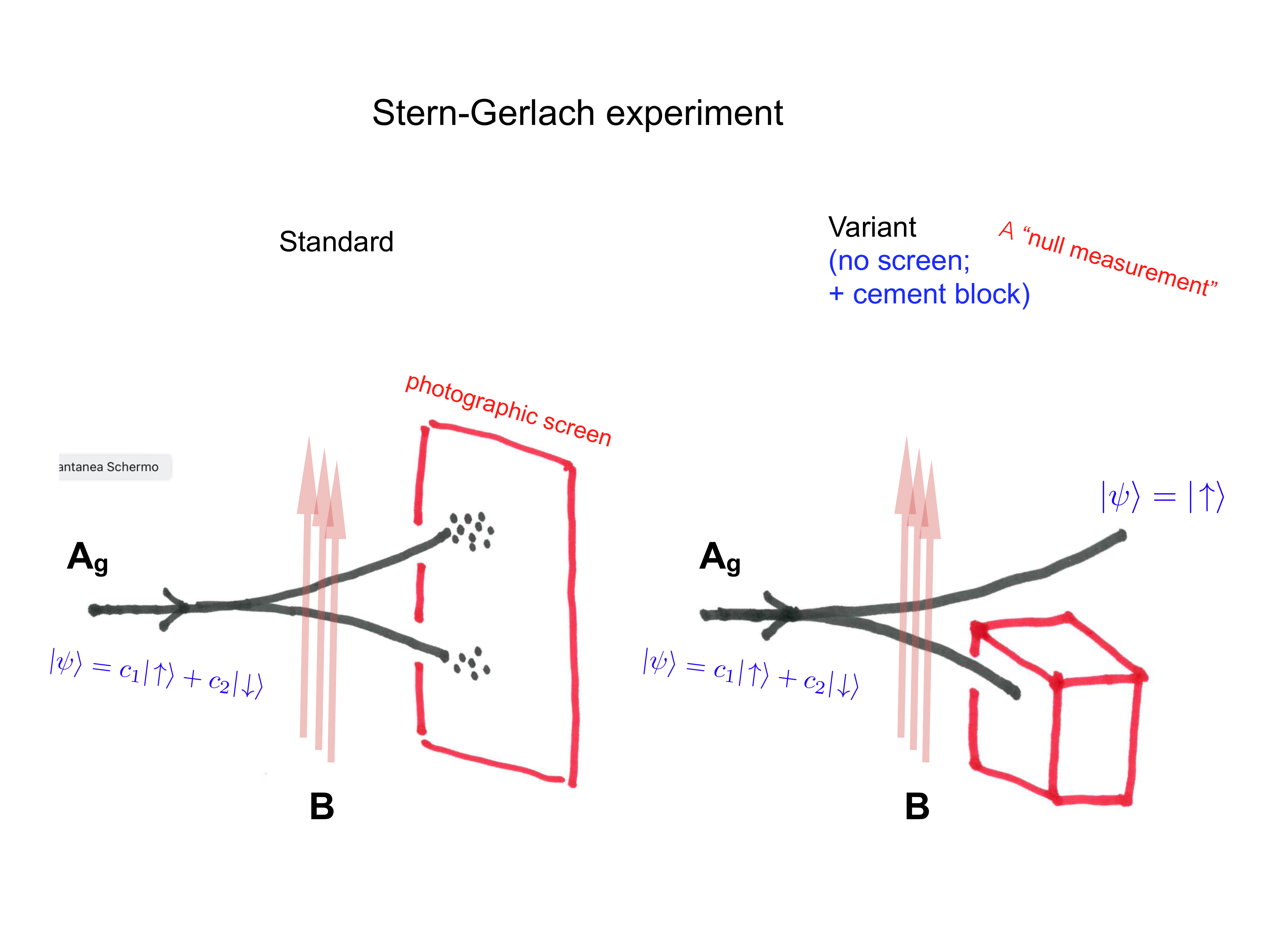}
\caption{\footnotesize   A variant (right) of the standard SG experiment (left).  A cement block stops the lower beam containing the spin-down atoms. 
No measurement is done (known as a ``null measurement"), but the atoms coming out of the region consist purely of spin-up state,  which can be used 
as the initial wave function  $|\psi(0)\ckt = |\!\uparrow\ckt$.  
}
\label{Variant}
\end{center}
\end{figure}

There are actually several types of measurements.  The first is known as {\it repeatable, or of  the first kind}:  it proceeds as 
\be      \left( \sum_n  c_n  | n \ckt  \right)    \otimes   |\Phi_0;    Env_0 \ckt     \longrightarrow    | m \ckt \otimes   |\Phi_m;  Env \ckt  \;, 
\ee
which corresponds to the expression often found in a QM textbook, 
\be      |\psi \ckt =     \left( \sum_n  c_n  | n \ckt  \right)   \longrightarrow   |m\ckt\;. 
\ee
In these experiments, the measuring device has acquired  and recorded the information about the microsystem under observation, but the latter
remains unscathed,  as a factorized, pure state.  
These do  represent  special class of measurements,  but they are  actually  neither  rare nor particularly difficult to realize.  A variant of  the Stern-Gerlach
experiment    (see Fig. \ref{Variant})    is a good example.

The very possibility of the repeatable measurements  is actually of fundamental importance for  QM.   They allow the preparation of a desired,  {\it  identical}, pure quantum state  $|\psi\ckt$   as the initial state, whenever  we wish.
This is important  because  in QM the experimental check of the theory is done  necessarily 
with many repeated, {\it  identical} experiments,   as QM predicts only the relative frequencies for various outcomes \footnote{Another, related fact is that   the atoms of the same kind, in their ground state, are all rigorously identical, in contrast to classical particles.   Such a characteristics of quantum mechanical world is at the basis of the  regularity of the macroscopic world, such as crystalline structures and extremely precise biological phenomena such as the reproduction.  }.

In contrast,  a macroscopic body made of,  say,  $10^{25}$   atoms and molecules  (e.g., an experimental device), can never be in an identical quantum state at  
different times or in different laboratories,  even though they might look identical at macroscopic scales. This is in fact  one of the  underlying reasons for the  state-vector reduction  (see the discussion above,   (\ref{JumpCor})).

The second type of the measurement may be simply termed {\it  non repeatable}.  This is  a general type of the measurement - state reduction,
as in (\ref{JumpCor}).   The result $F=f_m$ is recorded, but the original microsystem  itself (e.g., the electron in Tonomura's experiment)  simply gets lost,   in  $   |{\tilde \Phi}_m ; Env \ckt $.

There are also  third,   {\it  intermediate types} of  measurements.   The detection of particle tracks in Wilson, or  bubble chambers, or  in the silicon vertex detectors,  is an example of the measurement of this sort.  Note that in these experiments the microscopic particle interacts with the atoms in the devise, 
generates a local chain ionization cloud, but its information (the position or momentum) does not get lost completely, creating successive ionization clouds
 (particle tracks)  on its way.  This kind of processes  is particularly important in the momentum or energy measurement.

\subsection{Unitarity and linearity}

   The microscopic state  $\psi$   
\be      |\psi\ckt = \sum_n c_n  | n \ckt\;,      \label{stateU} 
   \ee
 if   {\it   left undisturbed},  evolves in time according to the Schr\"odinger equation, 
  \be  i \hbar  \frac{ d }{d t}   |\psi\ckt    = H\,  |\psi\ckt  \;, \label{Schr}
  \ee
 ($H$ being the Hamiltonian)   or as
   \be      |\psi\ckt    \longrightarrow    U\,   |\psi\ckt =     \sum_n   c_n  \,U  | n \ckt \;,    \qquad  U= e^{- i H t / \hbar}\;,    \quad    U^{\dagger}  U = {\mathbf 1}\;:   \label{Hin}
   \ee
   it is a linear, unitary evolution.   
   Unitarity  means that
   \be    \brc  \psi | \psi \ckt =   \sum_n  |c_n|^2  =1\;,  \label{normaU} 
   \ee
   i.e.,  the state  norms are conserved in time.    Linearity   means  (\ref{Hin}), i.e.,  superposition of different states  in (\ref{stateU})   continue to be  coherent superposition of the corresponding states.

 During the measurement process,  formally written as   (\ref{step111}),
     the time evolution might  still look unitary and linear.   
   However, as we have seen,  the wave functions associated with different terms in   (\ref{step111})  do not overlap in spacetime, so  it represents actually  a mixed state.  
   The coherent superposition of states is no longer there, see (\ref{lack}). 
   Most significantly,  the real time  evolution 
   of the system is the state-vector reduction,   (\ref{JumpCor}), meaning that
   {\it   linearity in the sense of (\ref{Hin})   is lost during the measurement. }

   On the other hand,  unitarity  is maintained \footnote{Note that linearity and unitarity  are two distinct concepts in quantum mechanics  (e.g., the momentum operator is linear but not unitary).    Dirac    noted   (Chap 27 of \cite{Dirac})  that the evolution operator must satisfy both, which are two independent requirements. They are both automatically met, once the choice  $U= e^{- i H t/\hbar}$ is made with a self-adjoint Hamiltonian operator,   and as long as the pure-state evolution  (\ref{Hin})   {\it  before}  the measurement,   is concerned.    
   }: 
   the sum of all possible outcomes, occurring at different measurements at different times,  adds up  to unity of  the total normalized frequencies,  corresponding to the  ``norm"   of the   state,  (\ref{step111}).
   In terms of the relative frequencies for different outcomes,    ${\cal P}_n$,    unitarity means
   \be    \sum_n   {\cal P}_n  = 1\;,    \qquad \sum_n  |c_n|^2 =1\;,  \label{obvious} 
\ee
even though different (sometimes the same, but repeated)  results $f_n$  refer to distinct measurements  {\it  made at different times.}

   Note that   this is  indeed  how experimentalists  view the meaning of unitarity.   Unitarity means that   $ {\cal N}_n / {\cal N}$  \footnote{ ${\cal N}_n$ is the number of times the experimentalist finds the result  $F=f_n$,  $n=1,2,\ldots$;   ${\cal N}$ is the total number of the measurements made.}, to be identified with the theoretical formula,  $   {\cal P}_n=    |c_n|^2$,
   should satisfy 
   \be \sum_n  {\cal N}_n / {\cal N} =1\;,  \label{trivial2}   \ee   
     which might look trivial.  However,  this contains an implicit, important   assumption  that 
 the  experimental device has no systematic bias, and registers all possible results $f_n$ with equal efficiency and with no losses.   
 Only in such an ideal measurement setting we can expect  that the experimental results $ {\cal N}_n / {\cal N}$ will approach  
  the  theoretical  prediction, ${\cal P}_n$,   in the limit of  {\it  large ${\cal N}$}  (e.g., Tonomura's experiment, Fig.~\ref{TonomuraExp}).

  It might be of some interest to compare the concept of unitarity in the measurement processes  as described here, with 
a more abstract  one,  in the conventional thinking based on Born's rule.   The latter means in fact
\be  \sum_n      P_n  = 1\;,\label{SumP}  
\ee
i.e., that the total    {\it probability}  for a single experiment   to give all possible results $f_n$,   is unity, and that it is conserved in time. 
The request  (\ref{SumP}), from the mathematical, logical point of view,  appears  quite  indispensable, and indeed has always been  considered as a sacrosanct principle in the traditional approach to quantum mechanics.

However,  once   $P_n$  is translated into the physically directly meaningful quantity, 
 ${\cal P}_n  ~(=P_n) $,
  the expected relative frequencies for various results $f_n$ which will  be 
 found in distinct measurements and  at different times,  unitarity as 
 an absolute sacred principle
may appear to lose part of the  {\it  aura}   surrounding it.

To conclude, one must be cautious  in applying  the concepts such as linearity or unitarity of the time  evolution, valid  in the context of (isolated)  pure states, to the measurement processes  where entanglement 
 between the microsystem and the measurement device as well as with the whole world,   
 plays an essential  role.  Their  consequences in these, complicated mixed states may well   (and indeed,  do) look differently from what one is accustomed to,  from  the 
study of  pure quantum states kept in isolation,   (\ref{Schr}), (\ref{Hin}).

\subsection{Measurements, emergence of classical physics, macroscopic quantum states, and models of wave-function collapses \label{MacroQM}}

The state after  a  measurement,  $  |{\tilde \Phi}_m ; Env \ckt $   (in (\ref{JumpCor}),  is an entangled,  {\it classical} state, with the recording of the measurement result.  The basic reason for that is the so-called envioronment-induced decoherence,  studied in \cite{Zurek}-\cite{Joos}.       
Emergence of classical physics in this context, is certainly one of the important  ingredients in the understanding of the quantum measurement processes  (the unique aftermath of each  measurement - the wave function collapse).   

In author's view,   however,  it is useful to consider the whole problem from a wider perspective of what may be called    ``great twin puzzles of physics today":   the general  quantum measurement problems  \cite{WheelerZ}-\cite{Bell}  on the one hand, and emergence of the classical mechanics from QM, on the other.   Even though these two classes of the problems are often discussed together \cite{Zurek}-\cite{Zurek2}, 
they are actually mostly independent issues   \footnote{This is so, even if  there are a few key questions which link the two classes of the problems: the classical behavior of the measurement devices  mentioned above is one of them. Another is the concept of the center-of-mass (CM) position and momentum of a macroscopic body, which requires necessarily a measurement to define them as the initial condition for studying the time evolution of the system.}. It is better to discuss them separately, and independently.   Here we are concerned with the first of them, the general measurement problems;    for the second  (how Newton's law emerges
from QM),  see a recent work,  \cite{KK2}.

For instance,  we know that sufficiently close to $T=0$, {\it  any} matter, in any state, is quantum mechanical. At
 $T=0$   any system is in the unique ground state   (Planck-Nernst law, or  the first law of thermodynamics).  

At  $T>0$ but sufficiently close to the absolute zero temperature, a macroscopic body can be in a coherent superposition,
\be   |\Psi\ckt =  c_1   |\Psi_1 \ckt  +    c_2   |\Psi_2  \ckt     \label{Macro}  
\ee
where $ |\Psi_1 \ckt  $ and  $|\Psi_2 \ckt  $ are macroscopically distinct states.   For experimental efforts to realize macroscopic (or mesoscopic) quantum states at very low temperatures,  
see   \cite{Leggett}-\cite{Brand}.

To talk about  the macroscopic quantum states as (\ref{Macro}) {\it   in connection with }  the measurement problems is, however,    slightly out of place.  Indeed,  
the scope of any  experimental device is opposite:    to  acquire  the information about the microscopic system  $\psi$ under consideration,  and to record it  in the form of a {\it   unique, classical}  state. 

When $|\Psi\ckt$  represents a system with an infinite number of degrees of freedom, there is an exception to the possibility of superposition of (for instance)
two, degenerate macroscopic states.   The system, at sufficiently low temperatures,  will choose one of them as its ground states.    It could be tempting to relate such a phenomena of spontaneous symmetry breaking  \cite{Nambu,Goldstone}   to the wave-function collapse.  For instance,  one might try to construct a dynamical model  \cite{Niew}  of spin $1/2$ particle, interacting with a doubly degenerate (2D)  Ising model ground states,  so as to mimic 
the state-vector reduction,  (\ref{JumpCor}). 

More generally,  the Ising model system may be replaced by  some classical ``pointer"  states;   the idea is to construct a model of nonunitary evolution
of the microsystem $\psi$ coupled to the pointer, to reproduce effectively  (\ref{JumpCor}), with correct relative frequencies for different outcomes \footnote{We will not try to make an exhaustive list of references here: some earlier ones are discussed in \cite{Bell};   many are cited in \cite{Niew} and in \cite{Froh}. There are also some talks presented in this conference.   }.

Whether or not such a  model of the wave-function collapse can be successfully constructed eventually, 
  the conclusion from our discussion in this section (and in \cite{KK}) is the following:    
  
   {\it    No dynamical models of wave-function collapse are needed. }  
   
In  our view,  ``wave-function collapse" is one of the worst misnomers  in the quantum mechanics discussions.  The words evoke in our mind a mysterious, nonlinear evolution which shrinks almost instantaneously whatever distribution present in $\psi$  before the measurement.   No such processes exist.  Every experimentalist knows what happens in each measurement: they are chain-ionizations and amlification,   hadronic cascades in a calorimeter, recording of the particle tracks, etc.  They represent all nonadiabatic, irreversible processes.  However complicated in detail,   they are all processes we understand well in principle,   in terms of  the standard theory of the fundamental interactions \cite{Weinberg}-\cite{GellMann}.  

 The state reduction 
 (\ref{JumpCor})  is just the way we perceive the fact that the result of each measurement   
  is necessarily one of the terms in (\ref{step111}).  The  wave functions of the states of the microsystem-device coupled system, 
 corresponding to different measurements,  have {\it  no common spacetime support}, hence the wave function apparently   - only apparently -   ``collapses".

\section{Three puzzles}

Before  concluding,  let us go through quickly three familiar puzzles and their resolution.

\subsection{Quantum jumps  versus Schr\"odinger's equation     \label{explain}}

A question   often debated is how to reconcile the  smooth  time evolution of the wave function described by the  Schr\"odinger equation  
with   sudden 
``quantum jumps",  occurring in radioactive nuclei  ($\alpha$ and $\beta$ decays),  or in  atoms in excited states, and so on.  

Let us consider an $\alpha$ decay from a metastable nucleus,  $X$, 
\be       {}^A_N \left( X  \right)    \Longrightarrow       {}^{A-4}_{N-2} \left( Y  \right)  + \alpha\;,
\ee
where $N$ is the atomic number and $A$ is the mass number.  The wave function for the 
system  may be written as  
\be        | \psi \ckt  =     | X \ckt    +    | Y \ckt  | \alpha  \ckt\;.   \label{nucleus}  \ee
 We do not bother to write the coefficients in front of  the two terms: 
their norms are not conserved due to the decay.
The first  ($ | X \ckt$)    corresponds to a bound state;  the second  ($ | Y \ckt  | \alpha  \ckt$)   an unbounded system.
   The interference between them is absent.  
      (\ref{nucleus}) is not the wave function of a pure state.  It is a mixed state. 
 
 The state being metastable,   
   its  energy has a small imaginary part, 
\be   E=  E_R -  i \, \frac{\Gamma}{2} \;, 
\ee
where $\Gamma$ represents the total decay rate per unit time  (or the level width), and  $\tau =   \hbar / \Gamma$ is  the  mean lifetime.   The wave function has a time dependence  (for instance, see \cite{KKGP}),
\be   {\tilde \psi}_X(t)  =     e^{- i E t / \hbar}  {\tilde \psi_X(0)  } =       e^{- i E_R t / \hbar}   e^{- \Gamma t /2 \hbar }   {\tilde \psi_X(0) }  \;.       \label{smooth}
\ee
Now how can one reconcile such a smooth time dependence with  sudden  quantum jumps, such as  $\alpha$ decay, a spontaneous emission of photons from an excited atom?  This sort of question, probably mixed up with a philosophical  confrontation between Heisenberg's  view  (based on the matrix mechanics, for  the transition elements)   and Schr\"odinger's  one  (with a smooth differential equations)  of quantum mechanics,   dominated the earlier debates on quantum mechanics \footnote{In fact, the two questions must be distinguished. The latter, more philosophical ``puzzle" was solved via  the proof of equivalence of the Hilbert spaces  $\ell_2$  and  $L^2$ by Schr\"odinger himself. See e.g.,  Tomonaga \cite{Tomonaga}:   it is not the subject of the discussion here.  }.

Actually,  from the very way the wave function $\tilde \psi$  for a metastable state is defined,  where  $\Gamma$ represents the total decay rates of all possible decay processes of the parent particle,  
it is quite clear  that   $\tilde \psi$   represents  an effective description of the metastable ``quantum state", in which the coarse-grained time dependence (decay events)  has been smoothed out. It does not have the same status as the wave function of a genuine  quantum state $\psi$.

In conclusion,  the conundrum of apparent  impossibility of  reconciling  the smooth time-dependent Schr\"odinger equation with quantum jumps,   appears to have been caused by  the confusion between  the concept of the true wave function  $\psi$  and
that of  an effective ``wave function" for a  metastable ``state", ${\tilde \psi}$.   

See \cite{KK} for more discussions.

\subsection{Schr\"odinger's  cat}

In the discussion involving Schr\"odinger's cat,   the initial ``wave function"    is  a combination  (\ref{nucleus}),  between the undecayed nucleus  $|X\ckt$
and the state after decay,   $ | Y \ckt  | \alpha  \ckt $.     To maximally simplify the discussion, 
let us eliminate altogether  the intermediate,  diabolic device which upon receipt of the $\alpha$ particle leads to the poisoning of the cat, and treat the cat  directly   as the measuring device, $|\Phi \ckt$.  When it detects the $\alpha$ particle, it dies;  when it does not, it remains alive. There are    
two terms in  the measurement process,   (\ref{step000}), (\ref{step111}),  which here reads 
 \bea   && | \psi \ckt \otimes |\Phi_0\ckt   =   \left(   | X \ckt    +    | Y \ckt  | \alpha  \ckt  \right)    \otimes |\Phi_0\ckt  \label{step00}  \\
 &  \longrightarrow   &      | X \ckt     \otimes |\Phi^{(alive)} \ckt     +      | Y \ckt  | \alpha  \ckt     \otimes    |\Phi^{(dead)} \ckt        \label{step11} \;
 \eea
 (we dropped also the rest of the world, $| Env \ckt $). 
Such a process appears  to lead to the superposition of the  dead and alive cat, which is certainly an unusual 
notion, difficult to conceive, to say the least. 

Actually, there are some abuse and/or misuse of concepts   in this argument, each of which individually invalidates it,  eliminating the notorious conundrum.

  The first is the fact, as seen in the last section, that  the ``state"  of the undecayed nucleus, $X$,  is not a pure state.   The  ``wave function"   ${\tilde \psi}_X$ describing it is not a proper wave function but an effective one, in which the coarse-grained time dependence has been averaged out.    Second,  the linear superposition 
$ |\psi \ckt  \sim   | X \ckt    +    | Y \ckt  | \alpha  \ckt $,   as  discussed around (\ref{nucleus}),    does not represent a pure
state,    but a mixed state:   $ | Y \ckt  | \alpha  \ckt $  represents the decay product of  
 $ | X \ckt $, unbounded and incapable of interfering with   the latter.

 Finally,  even putting aside these two  issues,   the process  (\ref{step00}),   (\ref{step11})   has clearly all the characteristics of a general  measurement process discussed in Sec.~\ref{measurement}.
As explained  there,    the 
 wave functions describing the two terms of (\ref{step11}) lack  overlapping spacetime supports.  
  There are no interferences between the terms involving the coupled system  involving  the microsystem  and the classical  measurement device (the cat here). The formula  (\ref{step11}) represents  a mixture.  
 
Equivalently,   and  perhaps more intuitively,    one can  use the notion of the wave-function collapse. 
The exact timing of the spontaneous  $\alpha$ decay cannot be predicted, as it reflects a quantum fluctuation. But the moment  an  $\alpha$ particle is emitted,  and the cat gets hit by it  (the instant of the measurement), the wave function collapses,  
 \be   | \psi \ckt \otimes |\Phi_0\ckt     \Longrightarrow      | Y \ckt  | \alpha  \ckt     \otimes    |\Phi^{(dead)} \ckt  \;.  
 \ee

 Recapitulating,  Schr\"odinger's cat paradox,   
was caused by improper use of concepts such as the superposition principle, unitary and linearity of evolution, and  by  mixing up the pure and mixed states   \footnote{ What happens actually is simple: as long as the nucleus has not decayed (no $\alpha$ particle emission) the cat remains alive. The exact timing of the $\alpha$ decay cannot be predicted. But the moment the $\alpha$ particle is emitted, and the cat gets hit,  it dies. 
That is  all.  
There are no problems in describing this process appropriately  in quantum mechanics, as seen above. }. 

Setting aside these arguments, 
{\it     the idea of superposition of dead and alive cat states,  represents in  itself a self-contradictory, impossible notion.}   A macroscopic superposition of states such as   (\ref{Macro}) or ({\ref{step11}),  
requires the body temperature  of the system close to $T=0$, whereas a living cat needs room (body-)  temperatures for its biological functions: it is necessarily a mixed state  \cite{KK2}.

\subsection{Quantum entanglement,  quantum nonlocality  and  EPR paradox \label{EPR}  }

Let us consider  now  the famous Einstein-Podolsky-Rosen setting (in Bohm's version) of a total spin $0$ system decaying into two spin $\tfrac{1}{2}$ particles
 flying away from each other.  
The (spin) wave function is given by 
\be     |\Psi_0\ckt =  \frac{1}{\sqrt 2}   \left(   |\!\uparrow\ckt |\!\downarrow \ckt  -   |\downarrow \ckt   |\!\uparrow\ckt       \right)      \label{Bohm}
\ee\!
where   $  |\!\uparrow\ckt$  ($ |\!\downarrow \ckt  $) represents the state of the 1st and 2nd  spins,  $s_{1, z}= \pm \tfrac{1}{2}$,  $s_{2, z}=\pm \tfrac{1}{2}$.    
As spin components of the single spins  do  not commute with the total spin  $  {\mathbbm S}_{tot}^2$,  e.g., 
\be   [   {\mathbbm S}_{tot}^2,   s_{1, i}]    \ne 0\;,\quad   [   {\mathbbm S}_{tot}^2,   s_{2, j}]    \ne 0\;,\qquad    i,j=x,y,z\;,
\ee
$s_{1\, z}, s_{2\, y} $, etc.,  are  fluctuating between $\pm \tfrac{1}{2}$ in the state $\Psi_0$.   Note that  $\Psi_0$ can be written in infinitely many different  ways, reflecting possible fluctuation modes of the subsystems, 
\be     |\Psi_0\ckt =   \frac{1}{\sqrt 2}   \left(   |+\ckt |- \ckt  -   |- \ckt   |+\ckt       \right)    =   \frac{1}{2}   \left[  \, |\!\uparrow\ckt (  |+\ckt -    |-\ckt )  -   (  |+\ckt -    |-\ckt )    |\!\uparrow\ckt   \,   \right]    = \ldots \;.\label{etcetc}  
\ee

It is of fundamental importance to realize that  such (correlated)  fluctuations of the two subsystems     -  {\bf  quantum entanglement} -    are present  however distant the subsystems may be, and  {\it  even if they are relatively spacelikely  separated. }
  Nothing in $\Psi_0$  tells  us otherwise.  Nothing in quantum mechanical law  says that  such entanglements  are 
present only when the subsystems are nearby, e.g.,  less than  $10^{-8}$ cm,  or   less than $10^{+28}$ cm.  Simply, 

 {\it  Quantum mechanics does not contain any fundamental parameter with the dimension of a length} \footnote{String theory, or quantum gravity, does have one, of the order of  the Planck length,  $\ell_P \sim10^{-33}$ cm.
We will not discuss  here either possible genuine modifications of quantum mechanics at such regimes, the issues of information-loss paradox and blackhole entropy, or  the consistent formulation of quantum gravity. Still, we do not take the view that either of them  (or   $\ell_P$)    might have any relevance to the measurement problems we are  discussing here.}.

In the EPR-Bohm experiment, this means that an experimental result at one arm,  say, $s_{1\,z}=  \tfrac{1}{2}$, might appear to   imply    ``instantaneously"  that the second experiment at the other arm would be in the state   $s_{2\,z}= - \tfrac{1}{2}$, even before 
it is actually performed, and however distant they may be.   This could sound paradoxical (sometimes the term ``quantum nonlocality"  is used).  The second experimentalist, not having access to the first experiment, might  have (should have?)  expected to find  the results  $s_{2\,z}= \pm \tfrac{1}{2}$,  a priori  with equal probabilities.   

 This kind of argument  has  led to the introduction of  the hidden-variable hypothesis  (see \cite{Bell} for discussions), although such a deduction was not really justified.  
We limit ourselves here to noting that this argument  had logical  flaws.  First, the contemporaneity of the two experiments is not really an issue: the experimentalists must just ensure that they are studying the two spin $\tfrac{1}{2}$ particles  from the same decay event -  the coincidence check. The exact time ordering, which depends on the 
reference system chosen,  cannot matter.      In any case, as the two subsystems are spacelikely separated, it is untrue that the second experimentalist would  know   instantaneously   the result of the first experiment,  or actually, even whether or not  the first measurement  has been indeed performed.     The information transmission is itself a dynamical process. 
Finally,  the two ``simultaneous" experiments would capture only those states of the two subsystems  
fluctuating according to the entangled wave function, (\ref{Bohm}),   (\ref{etcetc}). In conclusion,  for the second experimentalist the system presents itself as  a mixture with an  unknown density matrix.  {\it    He (she) simply does not know what to expect.  Therefore there was no paradox, whatsoever.}

It is true that  the first experimental result  $s_{1\,z}=  \tfrac{1}{2}$  {\it  does imply,  instantaneously}   (whatever it may mathematically  mean),  that the other spin is in  the state   $|{\downarrow}\ckt
= \tfrac{1}{\sqrt{2}} ( |+\ckt -    |-\ckt )$, as is seen from  the wave function  (\ref{Bohm}),  (\ref{etcetc}).  
This  quantum nonlocality, just a name for this particular aspect of quantum entanglement, is real.  {\it    It should, however,  not  be confused with the dynamical concept of locality (or causality)} 
\footnote{In author's view,  endless debates and confusions  about quantum mechanics have been  caused precisely  by such a confusion. }.       Quantum mechanics is perfectly consistent with causality and locality,  due to the fact that all the fundamental, elementary interactions are local interactions in spacetime  (see \cite{KK} and \cite{Peskin}).   No dynamical effects, including the information transmission,  propagate faster than the velocity  of light. 
  It is perhaps best to regard (\ref{Bohm})  simply as a particular kind of macroscopic  quantum-mechanical state.

It is  essential to realize that,   unless  the second measurement is done,     any components of the second spin  other than $s_{2\, z}$  are  still  fluctuating \footnote{This  shows that  the often mentioned 
  ``Bertlmann's socks" classical action-at-distance analogy,  is not valid.  Quantum non-locality is subtler,  and is  different  \cite{Bell}.  }.   Thus   if the
two Stern-Gerlach type experiments are performed at the two ends simultaneously, by using the magnets directed to generic directions ${\bf a}$ and ${\bf b}$,   
they would find,  experiment by experiment,   apparently random results, such as  $(s_{1\, a}, s_{2\,  b})  =    (\tfrac{1}{2}, \tfrac{1}{2}),  (\tfrac{1}{2}, - \tfrac{1}{2}),    (-\tfrac{1}{2}, -\tfrac{1}{2})$, etc.
But their fluctuation average is encoded in $\Psi_0$:
\be     \brc \Psi_0 | s_{a}  s_{b} | \Psi_0  \ckt =  -   \frac{1}{4}  \cos \theta_{a,b} =    -   \frac{1}{4} \,   {\bf a}\cdot {\bf b}\;.\label{QM}
\ee

A possible way to discriminate  between  any  alternative theory with hidden variables from quantum mechanics, 
has been mathematically  formulated, e.g., in the form of  the  Bell   \cite{Bell},      or CSHS   \cite{CHSH}    inequalities
for analogous,  polarization correlated photon pair  experiments.  Beautiful experiments by Aspect et. al.  (1981) \cite{Aspect} have subsequently demonstrated that,  whenever  the hidden-variable alternatives  and quantum mechanics
give discrepant predictions (for certain sets of polarizer axes), the experimental data confirm
quantum mechanics, disproving the former. 

See also  Chiao et. al. (1995) \cite{Chiao}  for a series of related experiments and discussions on quantum nonlocality.   See  \cite{KKGP,KK}  also  for comments on  and resolution to Mermin's conundrum  \cite{Mermin}  in  systems with entanglement among more than two particles.

~~~~

{\noindent {\bf A reflection} }

 As reviewed in Sec.~\ref{factent},   quantum correlations  (entanglement)  among the particles whose positions are  space-like-ly separated,  and hence are no longer capable to communicate with, or dynamically influence on, each other,   as the two spin $1/2$ particles in (\ref{Bohm}), (\ref{etcetc}),  are  quite a common, ordinary and ubiquitous   phenomenon in  QM.   What is not quite ordinary, is that this  ``quantum nonlocality"   does not represent any violation of dynamical causality or locality  (sometimes called Einstein causality).  Quantum mechanics is rigorously consistent with this principle, even though the proof of this fact requires working in the framework of relativistic quantum field theory, quantum theory of particles \cite{KK,Peskin}.     

Now  how can these two, apparently  contradicting  properties of quantum mechanics coexist peacefully within the same theory?   
The answer lies in another fundamental aspect of QM:  the wave function is not itself a measurable physical quantity (an observable). Only the relative frequencies for different outcomes (probabilities)  are predicted by it.

In hidden-variable models,  the statistical aspect of quantum mechanics is replaced by a classical statistical  (unknown)  distribution of the hidden parameters  $\{\lambda\}$,
complementing the standard description with the wave function.
  But once the value(s)  of  $\{\lambda\}$ at $t=0$ is (are) chosen,  all possible  imaginable future experimental outcomes are  determined uniquely
    (i.e., classical evolutions).
   
A  hidden-variable theory,  with a logical structure so sharply different from that of quantum mechanics, cannot reproduce {\it  all} of the predictions of the latter. 
It could, actually,  if one is willing to introduce nonlocal, noncausal evolutions.  This is seen clearly, e.g., in  the unphysical, noncausal  behavior of the trajectory $X(t)$ in Bohm's pilot-wave theory \cite{Bell},\cite{KKGP}   which is a particular type of  hidden-variable model, designed by construction  to reproduce all QM predictions. 

If we insist, instead, that a hidden-variable model must respect local causality, {\it then}  its predictions necessarily satisfy certain inequalities  such as those  
 in \cite{Bell}, \cite{CHSH}.     Bell and  Clauser et. al. have shown that {\it  some}  predictions  of QM  should  lie outside such a bound,  a fact  verified, in a dramatic experimental confirmation of QM  \cite{Aspect}.

\section{Conclusion}

Here are the answers to the ``quantum measurement problems"  enlisted in the beginning:

\begin{description}
  \item[(a)]      Wave function collapse?     
  
      {\it  No. See the Summary  A. and B.  after Eq. (\ref{JumpCor}), and the discussion at the end of Sec.~\ref{MacroQM}};
   \item[(b)]   Macroscopic superposition of states?  
   
       {\it  Yes (at temperatures close to $T=0$  and without environment-induced decoherence);  No, otherwise; }
  \item[(c)]   Forever branching manyworlds tree?   
  
   {\it   No;}
    \item[(d)]   Spontaneous collapse of micro state, of the experimental device, and of the world? 
    
      {\it  No;}  
      \item[(e)]   Wave function just a bookkeeping device?  
      
        {\it No;} 
        
        \item[(f)]   Quantum jumps vs Schr\"odinger equation?   
        
          {\it   Explained, see  Sec.~\ref{explain};}  
          
          \item[(g)]   Born's rule from the Schr\"odinger equation involving everything?  
          
            {\it  Yes, but only  effectively.   No, as a result a complicated nonlinear dynamical process.  See the Summary  A. and B.  after Eq. (\ref{JumpCor});}  
            
            \item[(h)]   Quntum nonlocality?   
            
              {\it  Yes,  see the discussions in Sec.~\ref{EPR};   }    
              
                Hidden variables?   
                
                 {\it No.    See the ``reflection" at the end of Sec.~\ref{EPR}.}
                 
                        \item[(i)]  Is it possible that the fundamental, complete theory of physics is incapable of unique prediction for a single measurement?    
                        
                         {\it   Yes.}    
\end{description}

The key observation of this work that the basic entities  (the degrees of freedom) of our world are various types of {\it  particles}, leads to the idea    that each measurement, at its core, is a spacetime, pointlike event.  The crux of the absence of coherent superposition of different terms in the measuring process (\ref{step111})  is, indeed,  the  consequent  {\it  lack of the overlapping  spacetime supports} in the associated wave functions, instant after measurement which is the process of  entanglement between the microsystem and the experimental device and the environment.
 It is in this sense the expression  (\ref{step111}) represents a mixed state.
   But it also implies that the aftermath of each measurement  corresponds to a single term of (\ref{step111}).    
This (also empirical) fact, is  known as the ``wave-function collapse",   or the state-vector reduction.

The discussions of Sec.~\ref{measurement}  have led to the diagonal density matrix, (\ref{diagonal}).  In particular,  
the information encoded  in the wave function    $|\psi\ckt    =    \sum_n  c_n   | n \ckt$  has been  transferred via entanglement with the experimental device and environment into the
relative frequencies for finding the experimental results $F=f_n$,      ${\cal P}_n = |c_n|^2$.  
Combined with the equality between   ${\cal P}_n = |c_n|^2$ (the relative frequencies) and $P_n = |c_n|^2$ (the probabilities) \cite{KK}  our results represent the first steps towards filling   in   the logical gaps in Born's rule, by  explaining the state-vector reduction, and pointing  towards a more natural interpretation of quantum mechanics.

   Our discussions  do   confirm the standard  Born rule as  a concise way of summarizing the predictions of quantum mechanics.  
But the fact that the latter  is formulated in terms of the  ``{\it  probability} certain result is found in an experiment,  which might  (or might not) be eventually  performed  - perhaps by an experimentalist with a Ph D \footnote{Borrowed from Bell's remarks, e.g.,   in ``Quantum mechanics for cosmologists" (reprinted in \cite{Bell}).}  -   has  always caused  confusion and conceptual difficulties
 \footnote{Every teacher of a quantum mechanics course knows about this curious feeling of guilt  he (or she) experiences, on the day of proclamation of the
``fundamental postulate"  (Born's rule).  }.

 {\it   In  ultimate analysis, the problem of  Born's rule   is  the fact that it 
presumes a human intervention.}

We instead take the quantum fluctuations described by the wave function
  as real,  and propose  that they represent the fundamental laws of quantum mechanics.   They are there, independently of any experiments, and in fact,  whether or not 
  we human beings are around  \footnote{From this point of view,  the notion of the ``wave function of the universe" makes perfect sense. 
   }.

What we propose here  is  a  slightly unconventional way of {\it  understanding} quantum mechanical laws.  Indeed, the solid textbook materials and the standard methods of analysis  
in quantum mechanics,  all remain unchallenged. 
After all, we know that quantum mechanics is a correct theory.

The new perspective, however,  suggests us to dispense with, in the discussion of quantum mechanics, 
certain familiar concepts  such as  {\it
 probability,  statistical
ensembles, Copenhagen interpretation,   etc., }
 which historically have guided us, but also caused  misunderstandings and  introduced clouds
in the discussion on quantum mechanics, for many years.

  Accepting the quantum fluctuations described by the wave function as real, 
we simply give up  pretending, or demanding,  that in a fundamental, complete theory describing Nature   the result of each single experiment should necessarily be predictable.

As  cosmologists tell us  today, 
all the known structures of our universe (including  galaxies, stars,  planets and ourselves) are believed to  have grown out of uncontrollable quantum density fluctuations at some stage of the inflationary universe.  In other words:

~~~

{\it God may not play dice,  but Nature apparently does. }

~~~

We believe that our new, clearer 
understanding the quantum mechanical laws of Nature,  has far-reaching consequences  and applications in all present and future science domains in physics and beyond, from  biology and chemistry,   quantum information and quantum computing,    and 
possibly, to the brain science.

\ack

We thank the organizers and participants of the  DICE2022  conference, Castiglioncello (Rosignano Marittimo, Tuscany),  Sept. 19-23, 2022, where this talk was presented,  for a stimulating atmosphere.    A special gratitude is reserved to Hans-Thomas Elze, for the  kind invitation and for many fruitful  discussions.  The author's  work is supported by an  INFN  special initiative grant,  GAST (Gauge and String Theories).

~~~~~

\end{document}